\documentclass[a4paper, fleqn,usenatbib]{mnras}

\usepackage[T1]{fontenc}
\usepackage{ae,aecompl}
\usepackage[colorinlistoftodos]{todonotes}
\usepackage{graphicx}	% Including figure files
\usepackage{amsmath}	% Advanced maths commands
\usepackage{amssymb}	% Extra maths symbols
\usepackage[normalem]{ulem}% For strikethrough

\newcommand{\alm}{a_{\ell m}}

\newcommand{\Cee}{\mathcal{C}}
\newcommand{\Cl}{\Cee_\ell}

\newcommand{\lmax}{\ell_{\rm max}}

\newcommand{\fsky}{f_{\rmn{sky}}}
\newcommand{\Nside}{N_{\rm side}}

\newcommand*{\expectation}[1]{\langle #1\rangle}

\newcommand*{\code}[1]{\textsc{#1}}

\newcommand{\healpix}{\code{healpix}}
\newcommand{\healpy}{\code{healpy}}
\newcommand{\camb}{\code{CAMB}}

\newcommand*{\satellite}[1]{\textit{#1}}
\newcommand{\COBE}{\satellite{COBE}}

\newcommand{\WMAP}{\satellite{WMAP}}
\newcommand{\Planck}{\satellite{Planck}}

\newcommand*{\Planckmap}[1]{\texttt{#1}}
\newcommand{\smica}{\Planckmap{SMICA}}
\newcommand{\common}{\Planckmap{Common}} % Common mask
\newcommand{\nilc}{\Planckmap{NILC}}
\newcommand{\sevem}{\Planckmap{SEVEM}}
\newcommand{\commander}{\Planckmap{Commander}}

\newcommand{\LCDM}{$\Lambda$CDM}

\newcommand{\PRIII}{\textit{PR3}}
\newcommand{\PRII}{\textit{PR2}}
\newcommand{\Forecast}{\textit{Forecast}}

\renewcommand*{\vec}[1]{\bmath{#1}}
\newcommand*{\unitvec}[1]{\vec{\hat{#1}}}

\newcommand*{\spinfunction}[2]{\,\vphantom{#1}_{#2}#1}

\newcommand*{\ssphYp}[3]{\spinfunction{Y}{+#1}_{#2 #3}}
\newcommand*{\ssphYm}[3]{\spinfunction{Y}{-#1}_{#2 #3}}

\def\lsim{\mathrel{\rlap{\lower4pt\hbox{\hskip1pt$\sim$}}
    \raise1pt\hbox{$<$}}}                % less than or approx. symbol
\def\gsim{\mathrel{\rlap{\lower4pt\hbox{\hskip1pt$\sim$}}
    \raise1pt\hbox{$>$}}}

\newcommand{\varT}{\textit{$\overline{T^2}$}}
\newcommand{\varP}{\textit{$\overline{Q^2 + U^2}$}}
\newcommand{\varPP}{\textit{$\overline{P^2}$}}

\newcommand{\iimag}{\rmn{i}}
\title[Hemispherical Variance Anomaly and $\tau$]{Hemispherical Variance Anomaly and Reionization Optical Depth}

\author[M. O'Dwyer, C.J. Copi, J. M. Nagy, C. B. Netterfield, J. Ruhl and G.D. Starkman]
{
M\'arcio O'Dwyer$^{1}$\thanks{E-mail: marcio.odwyer@case.edu}, 
Craig J. Copi$^{1}$, %\thanks{E-mail: cjc5@case.edu},
Johanna M. Nagy$^{2}$, %\thanks{E-mail: },\\ 
C. Barth Netterfield$^{2,3,4}$,
\newauthor
John Ruhl$^{1}$, %\thanks{E-mail: },
Glenn D. Starkman$^{1}$%\thanks{E-mail: glenn.starkman@case.edu}
\\
$^{1}$CERCA/Department of Physics/ISO, Case Western Reserve University, 10900 Euclid Ave.,
 Cleveland, OH 44106-7079, USA\\
$^{2}$Dunlap Institute for Astronomy \& Astrophysics, University of Toronto, 50 St George St, Toronto, ON, M5S 3H4,
Canada \\
$^{3}$Department of Astronomy and Astrophysics, University of Toronto, 50 St George St, Toronto, ON, M5S 3H4,
Canada \\
$^{4}$Department of Physics, University of Toronto, 60 St George Street, Toronto, ON M5S 3H4 Canada 
}

\date{Accepted XXX. Received YYY; in original form ZZZ}

\pubyear{2018}

\begin{document}
\label{firstpage}
\pagerange{\pageref{firstpage}--\pageref{lastpage}}
\maketitle

\begin{abstract}
Cosmic Microwave Background (CMB) full-sky temperature data show a
hemispherical asymmetry in power nearly aligned with the Ecliptic. In real
space, this anomaly can be quantified by the temperature variance in the
northern and southern Ecliptic hemispheres, with the northern
hemisphere displaying an anomalously low variance while the southern hemisphere
appears unremarkable (consistent with expectations from the best-fitting
theory, \LCDM). While this is a well-established result in temperature, the
low signal-to-noise ratio in current polarization data prevents a similar
comparison. Even though temperature and polarization are correlated, polarization realizations constrained by temperature data show that the lack of variance is not expected to be present in polarization data. Therefore, a natural way of testing whether the temperature result is a fluke is to measure the variance of CMB polarization components. In anticipation of future CMB experiments that will allow for high-precision large-scale polarization measurements, we study how the variance of polarization depends on \LCDM\ parameters' uncertainties by forecasting polarization maps with \Planck's MCMC chains. We find that, unlike temperature variance, polarization variance is noticeably sensitive to present uncertainties in cosmological parameters. This comes mainly from the current poor constraints on the reionization optical depth $\tau$ and the fact that $\tau$ drives variance at low multipoles. In this work we show how the variance of polarization maps generically depends on the cosmological parameters. We demonstrate how the improvement in the $\tau$ measurement seen between \Planck's two latest data releases results in a tighter constraint on polarization variance expectations. 
Finally, we consider even smaller uncertainties on $\tau$ and how more precise measurements of $\tau$ can drive the expectation for polarization variance in a hemisphere close to that of the  cosmic-variance-limited distribution.
\end{abstract}


%%%%%%%%%%%%%%%%%%%%%%%%%%%%%%%%%%%%%%%%%%%%%%%%%%

%%%%%%%%%%%%%%%%% BODY OF PAPER %%%%%%%%%%%%%%%%%%
%\listoftodos
\section{Introduction} \label{sec:intro}

Cosmic Microwave Background (CMB) temperature  anisotropies %are temperature fluctuations of black body radiation reminiscent of the Big Bang. These fluctuations 
have been measured over the full sky with increasing precision, beginning
with the Cosmic Background Explorer (\COBE) %, the first satellited designed to measure CMB anisotropies over the full sky, 
in the early 1990s, continuing with the Wilkinson Microwave Anisotropy Probe (\WMAP) in the 2000s and early 2010s, and culminating most recently in the \Planck\ satellite \citep{akrami2018planck} in the early to middle parts of this decade. These full-sky measurements from space have been crucially supplemented by numerous ground-based and balloon-borne experiments from the 1970s to today, and include, since the first detection of CMB polarization \citep{leitch2002measurement}, both temperature and polarization information. 

This wealth of  CMB data has been found to be largely in statistical agreement with expectations of the standard cosmological model, inflationary Lambda Cold Dark Matter (\LCDM). 
However, there are a few large-scale features of the CMB temperature fluctuations that are exceptionally unlikely according to \LCDM. 
These are the so-called large-scale anomalies of the CMB. 
Among these statistical anomalies \citep[for reviews of anomalies see][]{2010AdAst2010E..92C,2011ApJS..192...17B,ade2014planck,2015arXiv150607135P,2015arXiv151007929S} is the lack of power in the Ecliptic north temperature data at large scales, when compared with the all-sky power spectrum expectations. The power can be conveniently quantified in terms of the variance of the temperature field. The hemispherical variance anomaly was first reported by \citet{monteserin2008low} in the WMAP 3-year data \citep{2007ApJS..170..288H} and is closely related to the hemispherical power asymmetry, originally noticed \citep{Eriksen:2003db,hansen2004testing} in the 1-year \WMAP\ \citep{2003ApJ...583....1B} data, and  described as an asymmetry in power between hemispheres that are nearly optimally the northern and southern Ecliptic hemispheres. The power asymmetry can also be neatly described in terms of the variance by fitting dipoles to local variance maps of temperature data \citep{2014ApJ...784L..42A}. For more variance related studies see \citep{2007A&A...464..479B,2008JCAP...08..017L,lew2008hemispherical,cruz2011anomalous,2013JCAP...07..047G,ade2014planck}.

Even though CMB temperature anomalies have been studied extensively, especially since the 1-year WMAP data release in 2003, they remain unexplained.
One possibility, the so-called `fluke hypothesis', is that these anomalies are merely statistical fluctuations. 
A natural way of testing this hypothesis is to look for the same anomalies in other related cosmological observables -- perhaps the most obvious one being simply the CMB polarization components. 
Unfortunately current polarization data has too low a signal-to-noise ratio  on the relevant angular scales for anomaly detection \footnote{For \Planck's 2015 data products, modes as low as $\ell = 15$ need to be included for the hemispherical variance anomaly \citep{2015arXiv150607135P}. However, even lower multipoles are required depending on map, mask, and best-fitting cosmology. Additionally, anomalies such as the lack of correlation at large angles are present in multipoles down to $\ell=2$.}. 
This should begin to change with upcoming polarization experiments. \satellite{CLASS} \citep{essinger2014class} is mapping the polarization of CMB at large angular scales ($2 < \ell \lsim 200$) from a high-altitude site in the Atacama Desert in Chile. Covering 75 per cent of the sky, the data should be sufficient for a near sample-variance-limited measurement of the optical depth to reionization. CMB-S4 \citep{2016arXiv161002743A}, a possible future ground-based `stage 4' CMB survey to follow those currently being  deployed (`stage 3'), could consist of telescopes operating at the South Pole and the high Chilean Atacama plateau. \satellite{LiteBIRD} \citep{hazumi2012litebird} is a satellite mission that will survey the entire sky with a goal of detecting B-mode polarization.

One way to test for anomalies in polarization is through the prediction of well-defined test statistics in advance of upcoming experiments. This will obviate the concerns about \textit{a posteriori} statistics that currently plague the temperature anomalies. 
These test statistics would ideally be chosen on the basis of their ability to discriminate among predictive physical models for the anomalies, especially in the context of likely future observations.  
However, given the current absence of such models for most anomalies,  one may have to settle for statistics that can be used to test the fluke hypothesis. 
In \citet{o2017cmb} a statistic was proposed to test for the hemispherical variance anomaly in upcoming data.%CMB-S4 data. 
%The statistic was shown to reproduce the anomalous lack of variance in temperature, as seen in previous results in the literature. 
 The statistic was shown to detect the anomalous lack of variance in temperature in the ecliptic north, compared with \LCDM\ expectations, as seen in previous results in the literature.
It was also shown that, despite the expected correlation of temperature and polarization anisotropies in \LCDM, \LCDM\ CMB realizations constrained to have low  temperature variance in the north Ecliptic hemisphere did not display a corresponding anomalous lack of polarization variance in the Ecliptic north.
Therefore, variances of polarization maps are suitable statistics for testing the fluke hypothesis -- a low north-Ecliptic polarization variance might be expected in a physical model for low temperature variance, but would not be expected under the fluke hypothesis.

It should be noted that, thus far, analyses of test statistics, both to predict the variance distribution in polarization and to test it in temperature, have considered only best-fitting values of flat \LCDM\ parameters.
As shown in this work, neglecting parameter uncertainties is a reasonable assumption as far as the temperature goes; however,  polarization variance predictions prove to be much more sensitive to present uncertainties in cosmological parameters. 
This comes mainly from the current poor constraints on the reionization optical depth $\tau$. 
During the reionization period, high energy electrons inverse-Compton-scattered off CMB photons resulting in power injection predominantly at low multipoles ($\ell \lsim 30$). 
Consequently, variations in the value of $\tau$ result in significant variations in both the mode and width of the expected polarization variance distributions. 
It is therefore important to measure $\tau$ more precisely in order to increase the constraining power of the polarization variance test statistic.

In this work we show, using MCMC parameter chains from \Planck, how the variance of polarization maps generically depends on cosmological parameters. 
We demonstrate how the improvement in the $\tau$ measurement seen in \Planck's 2018 data release (\PRIII), when compared to the second (2015) release (\PRII), results in a tighter constraint on polarization variance expectations. 
Finally, by a scaling of \Planck's covariance matrix, we consider even smaller uncertainties on $\tau$. 
This improvement, which could result from a future experiment, conveys how more precise measurements of $\tau$ can drive the expectation for polarization variance in a hemisphere close to that of the  cosmic-variance-limited distribution.

The structure of the paper is as follows: in section \ref{sec:theory} the mathematical formalism of CMB anisotropies and their variance is presented; in section \ref{sec:tech} the technical aspects of the simulations performed and how the variance distributions were obtained are described; section \ref{sec:results} is dedicated to showing and discussing the results; finally, a summary and the conclusions are in section \ref{sec:conclusions}.

\section{Formalism}\label{sec:theory}
We represent CMB temperature anisotropies by $T(\unitvec{n})$, where the unit vector $\unitvec{n}$ denotes the direction of observation on the sky. 
Being a scalar quantity, $T(\unitvec{n})$ can be expanded in spherical harmonics,
\begin{equation}
	T(\unitvec{n}) = \sum_{\ell m} \alm^T Y_{\ell m}(\unitvec{n}).
\end{equation}
%where the fluctuation maps are represented by the coefficients $\alm^T$ in harmonic space. 
The polarization field, being a spin-2 quantity, can likewise be expressed in terms of spin-2 spherical harmonics. 
For the Stokes parameters of polarization we follow the convention of \citet{2005ApJ...622..759G},
\begin{eqnarray}
	Q(\unitvec{n}) & = & - \sum_{\ell m} \alm^E X_{1,\ell m}(\unitvec{n})
        + \iimag \alm^B X_{2, \ell m}(\unitvec{n}), \\
	U(\unitvec{n}) & = & - \sum_{\ell m} \alm^B X_{1,\ell m}(\unitvec{n})
        - \iimag \alm^E X_{2, \ell m}(\unitvec{n}), 
\end{eqnarray}
where the $\alm^E$ and $\alm^B$ are the standard $E$-mode and $B$-mode
coefficients and $X_{1, \ell m}$ and $X_{2, \ell m}$ are linear
combinations of spin-2 spherical harmonics defined by $X_{1, \ell m} \equiv
(\ssphYp{2}{\ell}{m} + \ssphYm{2}{\ell}{m})/2$ and $X_{2, \ell m} \equiv
(\ssphYp{2}{\ell}{m} - \ssphYm{2}{\ell}{m})/2$.

In \LCDM, the $\alm^X$ are  Gaussian-random variables of zero mean, hence the statistics of the fields can be fully described by their power spectra,
\begin{equation}
\label{eq:ClXY}
\Cl^{XY} = \expectation{\alm^X a^{*Y}_{\ell' m'}};\quad X,Y = \{T,E,B\},
\end{equation}
where the angle brackets denote an ensemble average, and the lack of dependence of $\Cl^{XY}$ on $m$ is a consequence of statistical isotropy. 
Given a set of cosmological parameters, the expected power spectra can be calculated. 
Maps can then be simulated by randomly drawing values of the $\alm^X$ coefficients according to the Gaussian statistics specified by \eqref{eq:ClXY}, and carrying out the appropriate sums. 

For a given temperature map, we define the temperature variance by \varT, where the bar denotes an average over the sky, either the full-sky or a specific region when partial coverage is considered. 
The Stokes parameters $Q$ and $U$ are individually coordinate dependent, so we consider instead the quantity $P \equiv Q + \iimag \, U$. The variance of polarization $\overline{P^2} = \varP$  is a rotationally invariant quantity.

\section{Technical details} \label{sec:tech}

Throughout the analysis and simulations we used the \healpix\footnote{See \url{https://healpix.sourceforge.net}} scheme to represent CMB maps and its \code{Python} wrapper \code{healpy}\footnote{See \url{https://healpy.readthedocs.io}} for spherical harmonic routines. 
We worked at a resolution of $\Nside =32$ and chose a maximum multipole of $\lmax = 64$. 
This resolution was chosen to optimize computational efficiency while retaining the relevant variance information.

\subsection{Data and masks} \label{sec:data}
To verify the low northern variance anomaly in temperature, we used \Planck's \smica\ temperature maps from the second and third data releases, \PRII\ and \PRIII. 
For each data release the appropriate \common\ temperature foreground mask was used to eliminate certain pixels before the variance calculation. 
The results for \nilc, \sevem\ and \commander\ were found to be consistent with \smica, and thus are omitted here for the sake of brevity. 

The data maps were degraded from the original resolution of $\Nside=2048$ to $\Nside=32$ by extracting the high-resolution set of $\alm$ up to $\ell = \lmax= 64$, deconvolving the high-resolution pixel window function and convolving the low-resolution pixel window function by appropriately re-scaling the $\alm$. (This is the method implemented in \citet{2015arXiv150607135P}.) 
To properly compare simulations with \smica, we compensated for smoothing by deconvolving the \smica\ anisotropies map with of a $5$ arcminute FWHM Gaussian beam. The mask on the other hand was degraded via \healpix's \texttt{ud\_grade} function. 
When degrading the mask, we matched the criteria used in the \Planck\ Release-2 analyses \citep{2015arXiv150607135P}, where all pixels with values less than $0.9$ are set to zero, and all others to one. 

We worked in the Ecliptic coordinate system to facilitate the separation of hemispheres in the data. 
Both data maps and masks were rotated from the Galactic projection to Ecliptic coordinates using \healpix's \texttt{rotate\_alm} post degrading. 
For masks, it is necessary to calculate the spherical harmonic coefficients from the real-space binary mask and, after rotation, transform back to real space. 
To correct for the resulting ringing while preserving the post-degrading relative sky coverage, pixels below $\approx 0.5$ were zeroed-out while pixels above this threshold were set to one.

To evaluate the impact of limited sky coverage on the variance distribution, we also considered a hypothetical mask that covers the portion of the Ecliptic northern hemisphere that can be seen from the Chilean Atacama site with foreground masked as in \Planck's \common\ mask. 
This sky coverage was calculated assuming that a ground base telescope at the aforementioned site covers Celestial latitudes from $60\degr$ South to $22.5\degr$ North, corresponding to Ecliptic northern $\fsky \approx 0.40$. 

\subsection{Simulations}\label{sec:sims}

To assess how the uncertainties on the measurement of \LCDM\ parameters influence the expected variance distribution for temperature and polarization, we considered the following Markov chain Monte Carlo (MCMC) chains:
\begin{enumerate}
	\item \Planck's 2018 TT,TE,EE + lowl + lowE chain. This chain (as well as the following) has near Gaussian posteriors with the following best-fitting values (and 68 per cent limits): $\Omega_b h^2 = 0.02236 \pm 0.00015$, $\Omega_c h^2 = 0.102 \pm 0.0014$, $100\theta_{\rm MC} = 1.04090 \pm 0.00031$, $\tau = 0.0544^{+ 0.0070}_{- 0.0081}$, $\ln(10^{10} A_s) = 3.045 \pm 0.016$ and $n_s = 0.9649 \pm 0.0044$. We denote this chain \PRIII.\footnote{
		Note that the nomenclature of the 2018 chains in \Planck's archives (\url{https://pla.esac.esa.int/\#cosmology}) is slightly different than the one used in \citet{aghanim2018planck}. In that publication, low-$\ell$ temperature data information is always assumed and therefore omitted from chain names, whereas in the archived 2018 chains the inclusion of this data is explicitly noted by the low-$\ell$ piece in the chain name.}
	\item \Planck's 2015 TT,TE,EE + lowP chain. The parameter intervals are $\Omega_b h^2 = 0.02225 \pm 0.00016$, $\Omega_c h^2 = 0.1198 \pm 0.0015$, $100\theta_{\rm MC} = 1.04077 \pm 0.00032$, $\tau = 0.079 \pm 0.017$, $\ln(10^{10} A_s) = 3.094 \pm 0.034$ and $n_s = 0.9645 \pm 0.0049$. We call this chain \PRII. Although there is no explicit `low-$\ell$' term in how \Planck\ labels the \PRII\ chain, it does include low-multipole temperature information. 
	\item A version of \PRIII\ where every (co)variance was divided by four. This artificial chain is used to showcase how a factor-of-two decrease in the uncertainty of $\tau$ impacts the expected variance distribution. 
	The whole chain is scaled so as to preserve the positive semi-definiteness of the covariance matrix. 
	We refer to this chain as \Forecast.
	As shown in Figure \ref{fig:cosmology-PR3-PR2}, this is sufficient for our purposes because the uncertainty in $\tau$ is the dominant effect in the difference between the cosmic-variance limited distribution of $\varPP$, and the distribution we infer from \PRII\ or \PRIII. 
\end{enumerate}
For each set of cosmological parameters in each chain we obtained the CMB power spectra $\Cl^{TT}$, $\Cl^{EE}$, $\Cl^{BB}$ and $\Cl^{TE}$ using the \code{Python} wrapper of \camb\footnote{See \url{https://camb.info/}}, \code{Python CAMB}\footnote{\url{https://camb.readthedocs.io/en/latest/}}. 
For each set of $\Cl$, we generated a random map of $T$, $Q$ and $U$ with \healpy, specifying $\Nside=32$, $\lmax=64$ and convolving the $\Nside=32$ pixel window function.
$\varT$ and $\varPP$ were calculated (with the relevant sky coverages/masks), enabling us to obtain their distributions for each chain.

We also produced simulated maps using the best-fitting power spectrum for each public data release of \Planck. 
For each release, the variance distributions were obtained by calculating the $\Cl$ assuming the best-fitting cosmology (with the best-fitting values showed in (i) and (ii)), drawing $50,\!000$ random maps, masking the maps and finding the variance over the unmasked region.

For further details on the cosmological parameters, see \citet{ade2016planck, aghanim2018planck}. 

\section{Results} \label{sec:results}
% eclipic north temperature variance for PR2 and PR3
\begin{figure}
\centering
	\includegraphics[width=\columnwidth]{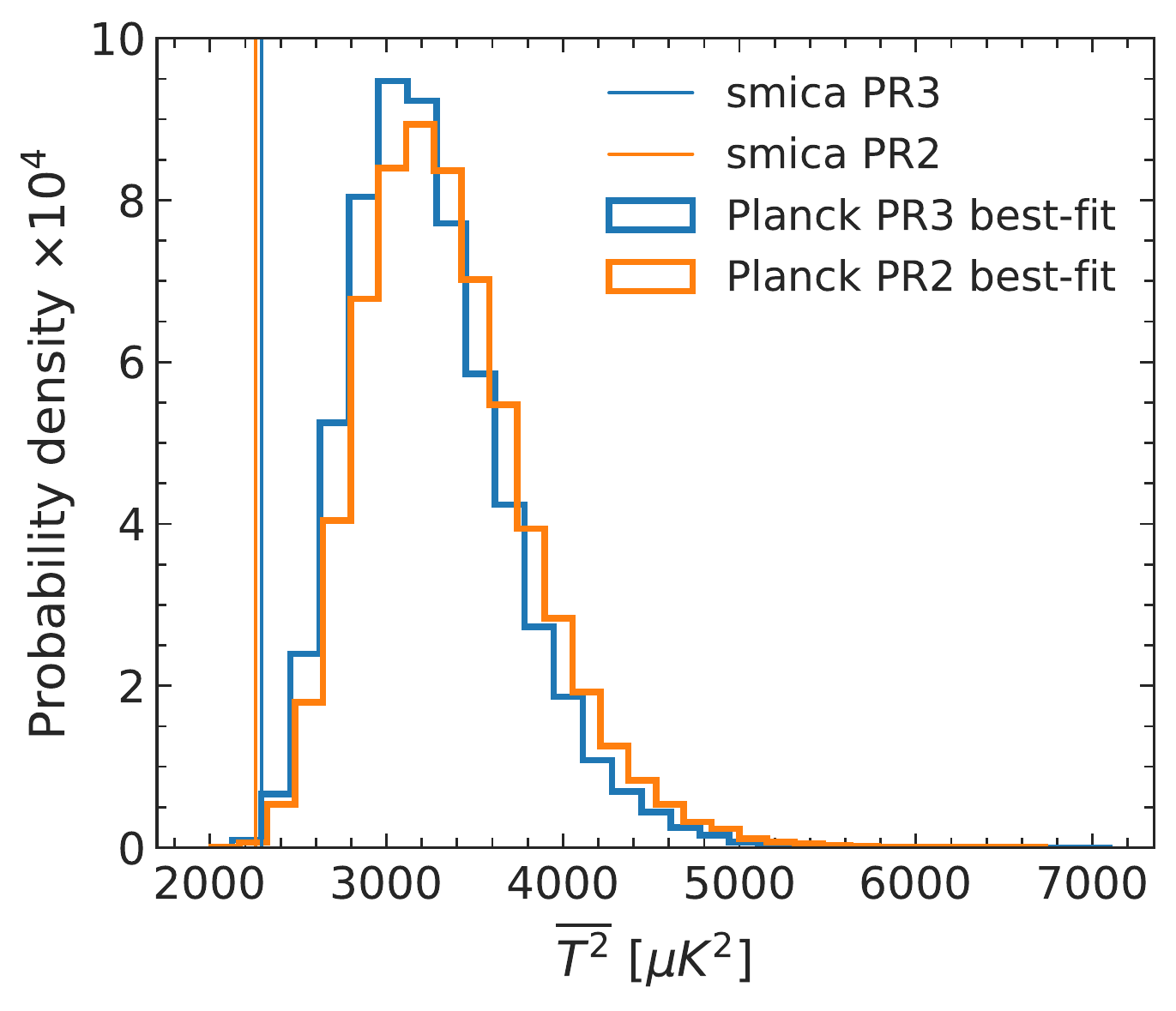}	
    \caption{Temperature variance distributions over pixels in the Ecliptic northern hemisphere with foreground masked according to \Planck's \common\ masks. 
    Vertical lines show the value obtained from temperature data utilizing the \smica\ map. 
    The distributions assume the best-fitting cosmology from \Planck's 2015 (orange) and 2018 (blue) data releases \citep{ade2016planck, aghanim2018planck}. 
    Data and mask of the appropriate release are used for the comparison. 
    Data variance has a $p$-value of 0.07 and 0.16 per cent for \Planck's 2015 and 2018 data and cosmology, respectively.}
    \label{fig:I-bf-PR2-PR3-EN-data}
    
\end{figure}

% var I and P for PR3 and PR2
\begin{figure*}
	\includegraphics[width=\textwidth]{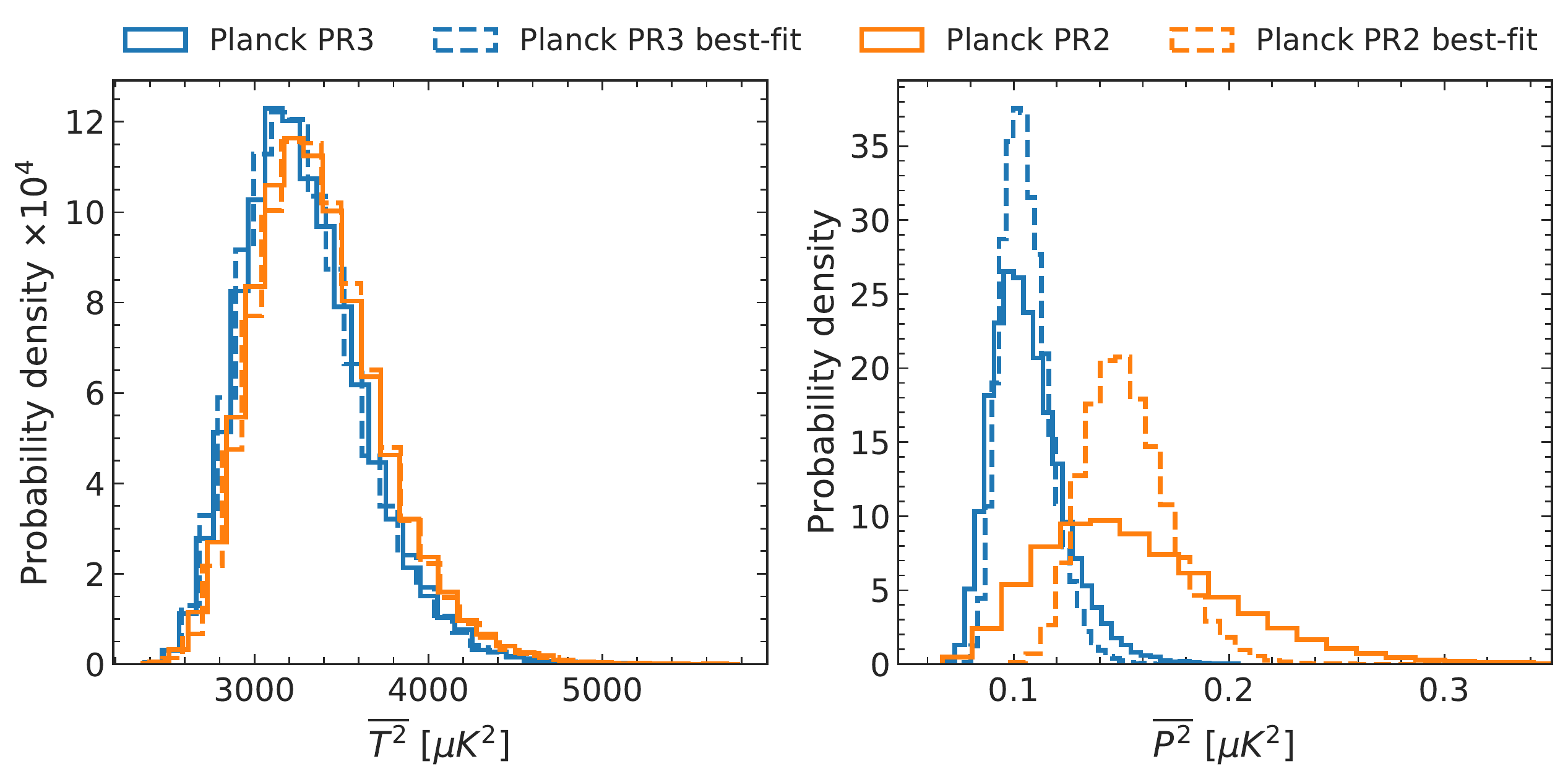}
    \caption{Temperature (left) and polarization (right) full-sky variances for \PRII\ (orange) and \PRIII\ (blue) cosmologies. 
    Distributions including uncertainties in cosmological parameters are given by solid lines, while the corresponding  distributions for the best-fitting cosmology are shown as dashed curves. 
    As can be seen from the left panel, temperature variance is essentially cosmic-variance limited, whereas the right panel shows that polarization variance is  sensitive to the cosmological error bars. 
    The comparison between \PRII\ and \PRIII\ polarization distributions highlights the change between data releases, as well as the better constraint in the expected full sky polarization variance.}
    \label{fig:I-P-fs-PR2-PR3}
\end{figure*}

Fig. \ref{fig:I-bf-PR2-PR3-EN-data} shows (vertical lines) the temperature variance in the northern Ecliptic hemisphere of \Planck's \smica\ maps from \PRII\ (orange) and \PRIII (blue) -- with foreground subtraction determined by \Planck's \common\ masks of the corresponding data release. These should be compared with the distributions of the temperature variance over the corresponding sky regions for the best-fitting cosmologies from \PRII\ and \PRIII.  
In the figure one can see how the variance in the Ecliptic north is anomalously low when compared to best-fitting-cosmology expectations. 
The likelihood of obtaining values as low as the ones seen in data is $0.04$ per cent for \PRII\ and $0.16$ per cent for \PRIII. 
The difference in p-value is mostly a result of the change in the \common\ masks from \PRII\ to \PRIII. 
Masking \smica's \PRIII\ map with the \PRII\ \common\ mask yields a p-value of $0.07$ per cent. 
The residual difference appears to be due to a slight shift in the best-fitting distribution.

In \citet{o2017cmb} predictions for the distribution of polarization variance were made  (for various portions of the sky) assuming the fiducial cosmological model. Here we study how uncertainties on \LCDM\ parameters impact the variance expectations. Since our goal is to simply assess this impact, we first consider full-sky simulations. 
The results for temperature (left panel) and polarization (right panel) for \PRII\ chains (orange) and \PRIII\ chains (blue) are shown in Fig. \ref{fig:I-P-fs-PR2-PR3}. For comparison, we display (dashed curves) the distribution assuming the corresponding \PRII\ and \PRIII\ best-fitting cosmologies.  

For temperature distributions, the chain and the best-fitting distributions are nearly identical for each release, and very similar from release to release. Therefore, temperature results obtained when assuming the fiducial model hold up when the uncertainties in cosmological parameters are included. However, it is clear from the second panel that this is not the case for polarization. 
The chain and best-fitting distributions differ noticeably, and there is a considerable shift in the mode and the width of the distributions between data releases.

% scatter plot matrix of variances in P - PR3 and PR2
\begin{figure*}
	\includegraphics[width=\textwidth]{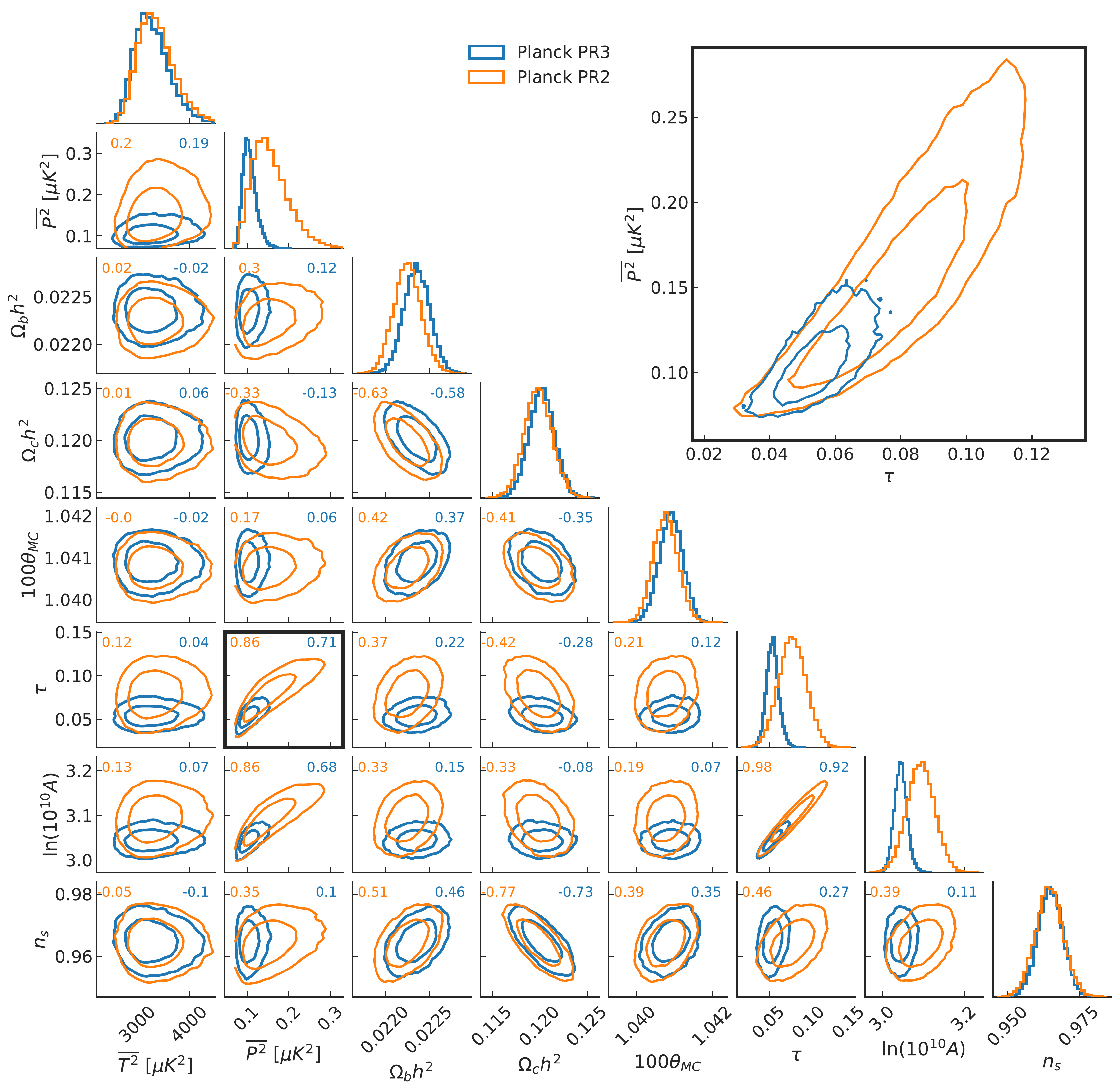}
    \caption{Confidence-curve matrix of temperature and polarization full-sky variances and flat \LCDM\ parameters for \PRII\ (orange) and \PRIII\ (blue) chains. The Pearson correlation coefficients for \PRII\ and \PRIII\ are shown as orange (top-left) and blue (top-right) text, respectively. From the first column it can be seen that temperature is mostly uncorrelated with cosmological parameters at this level of precision. The second column shows how polarization variance is correlated with the other quantities considered. Most of the uncertainty in \varPP\  comes from the fact that it is correlated with the reionization optical depth $\tau$, which in turn is poorly constrained by current measurements. The correlation comes from the fact that $\tau$ drives the reionization bump in the polarization power spectra. The correlation with $A$ is a consequence of the $\tau-A$ measurement degeneracy in \Planck\ data. The highlighted subplot, in evidence at the upper right corner of the figure, shows how the improvement in \Planck's measurement of $\tau$ impacts the expected variance distribution.}
    \label{fig:cosmology-PR3-PR2}
\end{figure*}

To gain insight into which cosmological parameters drive this behavior, we investigated  the correlations between the variances and each cosmological parameter. 
These are shown through a confidence-curve matrix presented in Fig. \ref{fig:cosmology-PR3-PR2}. 
\PRII\ results are again in orange and \PRIII\ in blue. 
The inner and outer curves represent boundaries of 68 and 95 per cent confidence levels, respectively. 
Examining the \varT\ correlations in the first column, we note that the temperature variance is nearly uncorrelated with every other quantity. 
The magnitudes of the Pearson correlation coefficients, $R$, between \varT\ and each \LCDM\ parameter (displayed in the figure in orange text on the top-left for \PRII\ and blue on the top-right for \PRIII) are $\leq 0.13$. 
The correlation  with $\varPP$ ($R \approx 0.2$) is as expected from $\Cl^{TE}$. 
This weak correlation between $\varT$ and \LCDM\ parameters is consistent with the similarity between chain and best-fitting distributions in the left panel of Fig. \ref{fig:I-P-fs-PR2-PR3}.

The stronger correlation of polarization variance with cosmological parameters is also  evident from the  second column of Fig. \ref{fig:cosmology-PR3-PR2} -- in the greater width of the chain distributions 
versus the best-fitting distributions, and in the shift in the best-fitting distributions between data releases. 
Most of the variation in chain parameters driving changes in \varPP\ comes from the optical depth $\tau$ (or consequently the amplitude $A$ of scalar fluctuations, which is nearly degenerate with $\tau$ in \Planck\ measurements). 
That is also as expected, since the reionization bump accounts for most of the power in polarization at $\ell \lsim 10$. 
This correlation is highlighted in the upper-right panel. 
It is also very clear from comparing \PRIII\ to \PRII\ how  tighter constraints on $\tau$ narrow the distribution of possible outcomes for \varPP. 
This improvement is a good example of how increasing the precision of the measurement of $\tau$ can enhance the constraining power of polarization variance, thus improving the ability to test the fluke hypothesis for the variance anomaly. 

% variace in P comparing bf, PR3 and forecast
\begin{figure}
	\includegraphics[width=\columnwidth]{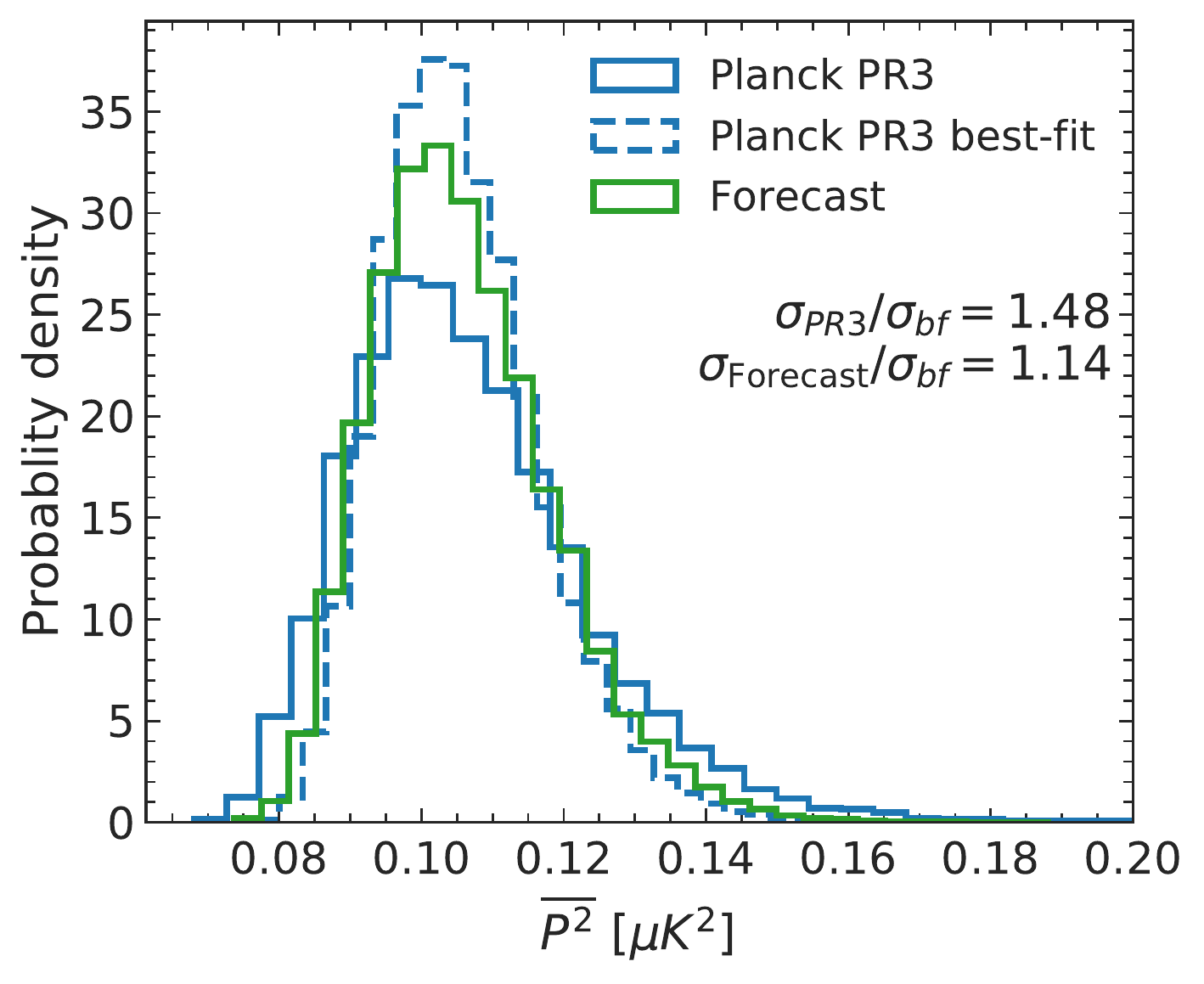}
    \caption{Full sky polarization variance comparing \PRIII\ (blue solid) and \Forecast\ (green solid) expectations with the best-fitting scenario (blue dashed). Distribution width is quantified by the standard deviation $\sigma$. The figure shows that the factor of two improvement in error bars from \PRIII\ to \Forecast\ result in narrowing of the variance distribution from 48 per cent to 14 per cent wider curves than the cosmic variance limited result.}
    \label{fig:P-PR3-forecast-bf}
\end{figure}

Although there has been a significant improvement from \PRII\ to \PRIII, the right panel of Fig. \ref{fig:I-P-fs-PR2-PR3} shows that the distribution of \varPP\ is still not cosmic-variance limited, in contrast to \varT. 
To illustrate how a further decrease in the uncertainty of $\tau$ would impact \varPP, we display the variance distribution for the \Forecast\ chain in Fig. \ref{fig:P-PR3-forecast-bf}. 
In the figure we compare the \PRIII\ curve (blue-solid) and the \Forecast\ curve (green-solid) to the curve obtained from best-fitting $\Cl$ (blue-dashed). 
%Using the standard deviation $\sigma$ as a metric, 
We find that the \PRIII\ distribution is about 50 per cent wider than the best-fitting distribution, as characterized by $\sigma_{PR3}/\sigma_{bf}=1.48$. 
For the \Forecast\ chain $\sigma_{\mathrm{Forecast}}/\sigma_{bf}=1.15$, clearly illustrating the impact on $\varPP$ of a factor of two improvement in $\sigma_{\tau}$.  
The impact of further improvements in $\sigma_{\tau}$ on \varPP\ are more difficult to quantify because the correlations of \varPP\ with other cosmological parameters become significant.

% variace in P for forecast- comparing masks
\begin{figure}
	\includegraphics[width=\columnwidth]{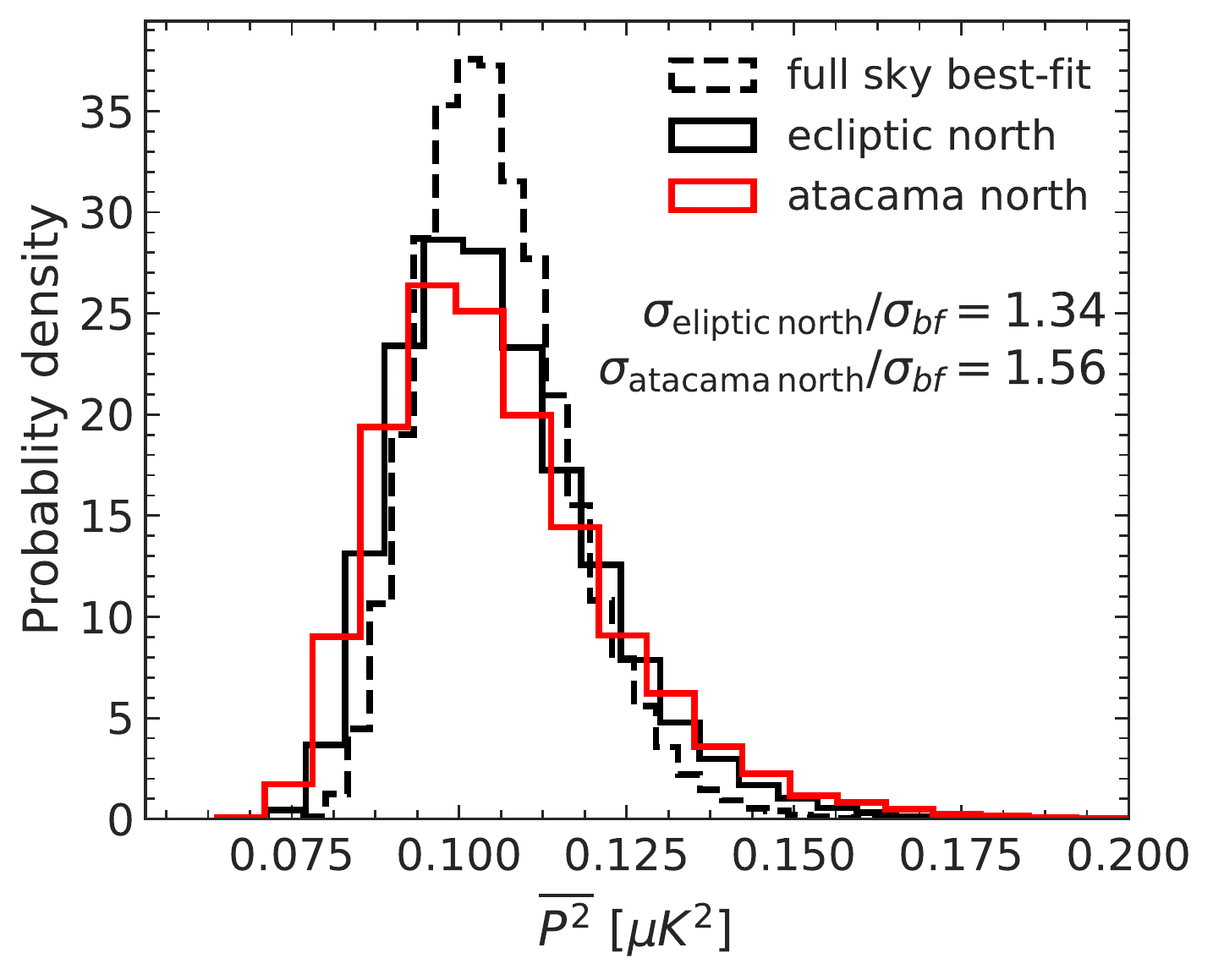}
    \caption{Polarization-variance distributions assuming \Forecast\ cosmology and error bars. We consider two coverage scenarios: pixels in the Ecliptic north (black solid) and the portion of the Ecliptic north seen from a telescope in the high Chilean Atacama plateau (red solid) with $\fsky \approx 0.40$ over the Ecliptic northern region. The black-dashed distribution displays the full-sky best-fitting result. This figure shows the impact of sky coverage on the width of the variance distribution compared with the best case scenario. While better-constraining $\tau$ should be a priority for testing the polarization variance anomaly, large sky-coverage is necessary to optimize the ability to test whether the variance anomaly in temperature carries over to polarization.}
    \label{fig:P-masks-forecast}
\end{figure}

Finally, Fig. \ref{fig:P-masks-forecast} shows the impact of sky coverage on $\varPP$, assuming the \Forecast\ chain. In it we compare full-sky coverage with two partial-sky scenarios: Ecliptic North coverage and the portion of the Ecliptic North seen from a ground-based telescope at an Atacama site. These are selected to demonstrate how different realistic sky coverages might affect the polarization-variance distribution. Both scenarios have had the foreground masked according to \Planck's 2018 \common\ mask, as described in Sec. \ref{sec:sims}. The full-sky best-fitting distribution of $\varPP$ is also shown for comparison.
We observe that, even though the greatest improvement would come from a better measurement of $\tau$, sky coverage also plays a role in how well we can predict $\varPP$. A large-sky-coverage polarization experiment would be ideal for testing large-scale anomalies.

\subsection{Additional discussion}

To avoid \textit{a posteriori} contamination of anomaly testing, it is necessary to design  statistics and make predictions for them before looking at data. 
In \citet{o2017cmb} it was shown that pixel variance is a suitable statistic to test in polarization for the low-northern-variance anomaly found in temperature. 
This utility assumed best-fitting cosmological parameters. 
In this work we showed how sensitive the predicted statistic is to the uncertainty in $\tau$. 
Consequently, we advocate for a more precise measurement of $\tau$ to enhance the discriminating power of our test statistics when large-angle, low-noise polarization data becomes available. 
Still, since \varPP\ and $\tau$ are so tightly correlated, the reader might wonder if they can be disentangled. 
In other words, to the extent that $\tau$ can be inferred from polarization variance, would anomalously low-variance data bias $\tau$ to a value that would in turn make the prediction statistic unsuitable for anomaly testing? 
If we look at the temperature result, it is clear that the anomaly comes from a conflict between data from a specific sky region (Ecliptic north) and best-fitting cosmology coming from full-sky measurements. 
Thus, it is natural to ask the same, or a similar question for polarization. Perhaps the question of the low variance anomaly could be recast as a comparison between $\tau$ measurements from different regions of the  sky. 

Additionally, temperature anisotropies at low multipoles are mostly due to the Sachs-Wolfe effect (with a subdominant contribution from the Integrated Sachs-Wolfe effect). 
Under the fluke hypothesis, it is this last-scattering-surface physics that is responsible for the low northern variance in temperature.  
A polarization-based test of the fluke hypothesis rests on the low-$\ell$ polarization signal also coming from  last-scattering-surface physics. The variance due to $\tau$, reflecting a secondary anisotropy source, should be considered as noise. 
Therefore, it might be desirable to engineer a variance statistic that `denoises' the contribution from $\tau$. One natural way to go about this is to simply exclude data at scales where reionization dominates polarization power. However, one must ensure that the anomalous temperature feature is still present under the same conditions. The variance dipole anisotropy is present  in \Planck's 2015 data even when $\ell<15$ are excluded \cite[Sec. 6.1]{2015arXiv150607135P}. However, we have found that \Planck\ 2018 data does not have anomalously low temperature variance in the Ecliptic north when multipoles below $\ell<10$ are excluded. Thus one should be cautious when filtering out low $\ell$ modes.

\section{Conclusions} \label{sec:conclusions}

The Cosmic Microwave Background temperature anisotropies exhibit an anomalous lack of variance in the Ecliptic northern hemisphere. 
To investigate whether or not this is a statistical accident (the `fluke hypothesis'), one can examine future polarization data, which in \LCDM\ is expected not to exhibit an analogous anomaly \citep{o2017cmb}. 
We have predicted the probability distribution of the variance of CMB polarization maps considering uncertainties on flat-\LCDM\ parameters. 
To do so we considered \Planck's 2018 TT,TE,EE + lowl + lowE and 2015 TT,TE,EE + lowP chains. 
We found that the expected distributions are noticeably wider than when the cosmological parameters are fixed to their best-fitting 2015 values.
However, there has been a considerable improvement in 2018 results, compared to 2015. 
By looking at the correlation of variance with \LCDM\ parameters, we find that the change is a result of \Planck's enhanced measurement of the reionization optical depth $\tau$. 
This is an intuitive result since $\tau$ is currently poorly constrained while also being the main source of low-$\ell$ power in the polarization power spectra.

We then considered a new chain, which was obtained by scaling the covariance matrix of \Planck's 2018 chain by a factor of four, effectively reducing the standard deviations in cosmological parameters by a factor of two. Given the $\tau$-variance correlation, and the fact that $\tau$ is poorly constrained compared to other \LCDM\ parameters, this chain can be used to approximately analyze how varying constrains on $\tau$ impacts polarization variance. Utilizing this chain we showed how the factor of two improvement in measuring $\tau$ resulted in a roughly 30 per cent narrower probability distribution for polarization variance. The resulting distribution is only 14 per cent wider than the cosmic variance limited result. Therefore, it is clear that a more precise measurement of the reionization optical depth would improve the power of measurements of  polarization variance to test the predictions of \LCDM,
and compare them to the hypothesis that whatever is causing the lack of variance in temperature would correspondingly affect polarization variance.

We highlight that although the error in $\tau$ is the main driver of uncertainty in the expected variance distribution, sky coverage also plays a big role in the distribution's width. We showed how the width of the variance distribution depends on  foreground-subtracted Ecliptic northern sky coverage. 
We compared what could be learned from  the portion of the Ecliptic north seen from a telescope in the high Chilean Atacama plateau (as expected for the \satellite{CLASS} experiment) to the best-case scenario of full sky coverage and best-fitting parameters. 

In this work we assumed that the value of $\tau$ was the same over the entire sky. In a future work, it would be interesting to study how a local dependence of $\tau$ could be related to a possible low variance anomaly in polarization. In that vein, a particularly interesting experiment would have large enough coverage and precision that the value of $\tau$ could be inferred independently from different parts of the sky, so that comparisons between regions are possible. Ideally these independent determinations would include the northern and southern ecliptic hemispheres. A future challenge may be to disentangle consequences of patchy reionization from the effects of the north-south temperature variance anomaly carrying over into polarization.

We acknowledge that if there is a mechanism suppressing variance at the last scattering surface, then polarization variance due to reionization should be considered noise for low-variance anomaly testing. From that point of view, the work in this article shows how this noise impacts the variance statistic as a function of $\tau$. To test for this scenario, one would have to somehow strip this contribution off of polarization variance. The simplest way being just not considering multipoles beyond a certain threshold determined by the reionization bump contribution to the power spectra. However, it is unclear whether or not temperature variance is robustly anomalous in that case. This line of thought is reserved for a future work.

After the completion of the analysis contained in this work,  the Planck Team released their 2018 Isotropy and Statistics paper \citep{Akrami:2019bkn} in which they identify a hemispherical variance anomaly in E-mode polarization, which they characterize as a dipole. Of the full-sky E-mode maps prepared using Planck's four standard  diffuse component separation methods, two (Commander and SEVEM) present amplitudes of that dipole that are particularly anomalously high (p-value $<1\%$), while for two (Commander and SMICA) the direction is particularly anomalously parallel ($p<1\%$) to the equivalent direction for temperature.  According to the Planck Team, noise, especially anisotropic noise, remains a  concern and they look to ``future data sets'' to provide ``additional insight''.  This supports the need for improved determination of $\tau$ to clarify the interpretation of those future data sets.
\section*{Acknowledgements}

We thank Joshua Ziegler for early work that suggested the need to explore the effects of uncertainties in $\tau$ on the variance anomaly.
GDS and MO’D are partially supported by Department of
Energy grant DE-SC0009946 to the particle astrophysics
theory group at CWRU. Some of the results in this paper have been derived using the
\healpix\ \citep{2005ApJ...622..759G} package. 
This work made use
of the High Performance Computing Resource in the Core
Facility for Advanced Research Computing at Case Western
Reserve University.

%%%%%%%%%%%%%%%%%%%%%%%%%%%%%%%%%%%%%%%%%%%%%%%%%%



\bibliographystyle{mnras}
\bibliography{tau_vs_var} % if your bibtex file is called example.bib

%%%%%%%%%%%%%%%%%%%%%%%%%%%%%%%%%%%%%%%%%%%%%%%%%%

%%%%%%%%%%%%%%%%% APPENDICES %%%%%%%%%%%%%%%%%%%%%

\appendix

%%%%%%%%%%%%%%%%%%%%%%%%%%%%%%%%%%%%%%%%%%%%%%%%%%

% Don't change these lines
\bsp	% typesetting comment
\label{lastpage}
\end{document}